# Relationship between exponent of power-law distributions and exponent of cumulative distributions[*]

GUO Jin-Li

(Business School, University of Shanghai for Science and Technology, Shanghai 200093, China)

**Abstract:** We commented on Ref.[Andrade J S, Herrmann H J, Andrade R F S, et al. Phys. Rev. Lett. 94, 018702 (2005)] and corrected the approach to estimate the degree distribution of the Apollonian network. However, after reading our manuscript, Herrmann indicated that it was due to a small typographic error and Herrmann et al. published Ref. [Andrade J S, Herrmann H J, Andrade R F S, et al. Phys. Rev. Lett. 102, 079901 (2009)]. In this paper, the relationship between an exponent of power-law distributions and the exponent of cumulative distributions is studied. For power-law distribution with geometrically growing domain, we prove that its exponent is equal to the exponent of its cumulative distribution. We carried out numerical simulations and obtain results that are in good agreement with the theoretical analysis.

**Key words:** Complex network; Apollonian network; Power-law distribution; Exponent.

Herrmann et al. calculated an exponent of the degree distribution of an Apollonian network by utilizing the method that the exponent of a power-law distribution equals the exponent of its cumulative distribution plus one. We commented on Ref.[1], pointed out an error of the exponent of the Apollonian network and corrected the approach to estimate the exponent of its degree distribution. Editors of PRL send our comment to Herrmann for a review. However, after reading our manuscript, Herrmann indicated that it was due to a small typographic error and Herrmann et al. published Ref. [2]. We hope that scholars study this problem because the cited rate of Ref.[1] is very high and errors of many papers were caused by this paper. In our paper, we study the relationship between the exponent of a power-law distribution and the exponent of its cumulative distribution. For power-law distribution with geometrically growing domain, we prove that its exponent equals the exponent of its cumulative distribution. We carried out numerical simulations and obtain results that are in good agreement with the theoretical analysis. We question that the exponent of degree distribution is estimated by the exponent of cumulative degree distribution in a study of complex networks.

** Project 70871082 supported by the National Natural Science Foundation of China (Grant No 70871082) and Project S30504 supported by the Shanghai Leading Academic Discipline Project (Grant No S30504).
Corresponding author. *E-mail address*:  phd5816@163.com.

Following on concepts of random variables taken from Refs.[3,4]. A set of all possible outcome of an experiment is known as the sample space of the experiment and is denoted by $S$. Real-valued functions defined on the sample space are known as random variables. We say that $Y$ is a continuous random variable if there exists a nonnegative function $f(x)$, defined for all real $x \in (-\infty, +\infty)$, such that, $P\{Y < +\infty\} = \int_{-\infty}^{+\infty} f(x)dx = 1$, and

$$F(x) = P\{Y < x\} = \int_{-\infty}^{x} f(x)dx \qquad (1)$$

The function $f(x)$ is called the probability density function of the random variable $Y$. The function $F(x)$ is called the distribution function of the random variable $Y$. The function $\overline{F}(x) = 1 - F(x)$ is called a cumulative distribution function[1] or a complementary distribution function of the random variable $Y$.

For sufficiently small fixed $\Delta x > 0$,

$$P\{Y = x\} \approx P\{x \leq Y < x + \Delta x\} = \int_{x}^{x+\Delta x} f(x)dx \approx f(x)\Delta x \qquad (2)$$

From Eq.(2), we see that $f(x)$ is a measure of how likely it is that the random variable $Y$ will be near $x$.

A random variable that can take on at most a countable number of possible values is said to be discrete. For a discrete random variable $\xi$, we define the probability mass function $p(x)$ of $\xi$ by

$$p(x_i) = P\{\xi = x_i\} = p_i \geq 0, i = 1, 2, \cdots \qquad (3)$$

and $\sum_{i} p_i = 1$. Then the cumulative distribution function of the discrete random variable $\xi$ is

---

[1] $\overline{F}(x)$ is said to be a cumulative distribution function In Ref.[6], it is called a complementary distribution function in Ref.[7].

obtained from Equation (3) as follows

$$\overline{F}(x) = \sum_{x_i \geq x} p_i \tag{4}$$

**1. Continuous power-law distribution**

Continuous random variable $Y$ is said to obey a Pareto distribution or a power-law distribution if its probability density function is given by

$$f(x) = \begin{cases} ra^r \dfrac{1}{x^{r+1}}, & \text{if } x \geq a \\ 0, & \text{if } x < a \end{cases} \tag{5}$$

where, $a > 0, r > 0$. $\beta = r + 1$ is said to be a exponent of this distribution. The cumulative distribution of the random variable $Y$ is given by

$$\overline{F}(x) = \dfrac{a^r}{x^r} \tag{6}$$

From Eq. (5) and Eq. (6), for the continuous random variable $Y$, the exponent of the power-law distribution equals the exponent of the cumulative distribution function plus one.

**2. Discrete power-law distribution**

Non-negative discrete random variable $\xi$ is said to obey a power-law distribution if the probability mass function is given by

$$p(x_i) = P\{\xi = x_i\} = ra^r \dfrac{1}{x_i^{r+1}}, \quad i = 1, 2, \cdots \tag{7}$$

and $\sum_{i=1}^{+\infty} ra^r \dfrac{1}{x_i^{r+1}} = 1$. The cumulative distribution of the random variable $\xi$ is given by

$$\overline{F}(x) = \sum_{x_j \geq x} ra^r \dfrac{1}{x_j^{r+1}} \tag{8}$$

This is a step function. In applications, one concerns $x = x_i, i = 1, 2, \cdots$. From Equation (7), a domain of the power-law distribution is the same as a range of the random variable $\xi$. We assume that the domain of the cumulative distribution of the discrete random variable $\xi$ is the same as the range of this random variable.

**2.1 Discrete random variable with continuously increasing range**

Let $m > 1$, we assume that non-negative random variable $\xi$ obeys the power-law distribution with range $\{m, m+1, m+2, \cdots\}$ and its probability mass function is given by

$$p(k) = P\{\xi = k\} = ra^r \frac{1}{k^{r+1}}, \quad k = m, m+1, m+2, \cdots \qquad (9)$$

Then its cumulative distribution function is given by

$$\overline{F}(k) = ra^r \sum_{i=k} \frac{1}{i^{r+1}} \qquad (10)$$

For an arbitrary given $r > 0$, the sum in Equation (10) is very difficulty, it may be approximated by

$$\overline{F}(k) = ra^r \sum_{i=k} \frac{1}{i^{r+1}} \approx ra^r \int_k^{+\infty} \frac{1}{x^{r+1}} dx = \frac{a^r}{k^r} \qquad (11)$$

The approximation in Equation (11) makes use of an integral definition. Since the range of the random variable $\xi$ is continuously increasing, Equation (11) implies factor $\Delta x = i - (i-1) = 1$. For discrete random variable with continuously increasing range, from Eq. (9) and Eq. (11), we know that the exponent of the power-law distribution is equal to the exponent of its cumulative distribution plus one. This conclusion is widely used in the study of complex networks. For discrete random variable with geometrically increasing range, although it is also widely used in the study of complex networks, this conclusion is not justified.

**2.2 Discrete random variable with geometrically increasing range**

Let $m > 1, c > 0$, we assume that non-negative discrete random variable $\xi$ obeys a power-law distribution with range $\{c, cm, cm^2, cm^3, \cdots\}$ and its probability mass function is given by

$$p(k) = P\{\xi = k\} = ra^r \frac{1}{k^{r+1}}, \quad k = c, cm, cm^2, cm^3, \cdots \qquad (12)$$

Then its cumulative distribution function is given by

$$\overline{F}(k) = ra^r \sum_{i=l} \frac{1}{(cm^i)^{r+1}} = \frac{ra^r m^{r+1}}{(m^{r+1}-1)k^{r+1}}, \qquad k = cm^l, l = 0,1,2,\cdots \quad (13)$$

For discrete random variable with geometrically increasing range, from Eq. (12) and Eq. (13), the exponent of power-law distribution is not equal to the exponent of its cumulative distribution plus one, but its exponent is equal to the exponent of its cumulative distribution.

In double logarithmic coordinates, numerical simulations of the power-law distribution with geometrically increasing range and its cumulative distribution are in Figure 1. The exponent of the power-law distribution is the same as the exponent of its cumulative distribution, see Fig. 1.

Expectation of random variable $\xi$ is given by

$$E\xi = ra^r \sum_{i=0} \frac{1}{(cm^i)^r} = \frac{ra^r m^r}{(m^r-1)c^r} \quad (14)$$

If $r > 0$, the expectation is finite.

Herrmann et al. obtained exponent $\ln 3/\ln 2$ of the cumulative degree distribution of the Apollonian network [1]. However, the degree of this network can be regarded as the random variable with range $\{3, 3\times 2, 3\times 2^2, \cdots, 3\times 2^{t-1}, \cdots\}$. Thus, the exponent of the degree distribution of the Apollonian network should be $\ln 3/\ln 2$. From Ref.[5], we know that the degree distribution of the Apollonian network is given by

$$p(k) = \frac{2}{3}\left(\frac{3}{k}\right)^{\frac{\ln 3}{\ln 2}}, \quad k = 3, 3\times 2, 3\times 2^2, 3\times 2^3, \cdots \quad (15)$$

By comparing Eq. (12) with Eq. (15), we have

$$r = \frac{\ln 3}{\ln 2} - 1, \quad c = 3, \quad m = 2, \quad ra^r = \frac{2}{3} \times 3^{\frac{\ln 3}{\ln 2}}.$$

From Equation (13), we know that the degree of the cumulative distribution of the Apollonian network is given by

$$\overline{F}(k) = \left(\frac{3}{k}\right)^{\frac{\ln 3}{\ln 2}}, \qquad k = cm^l, l = 0,1,2,\cdots \quad (16)$$

From Equation (14), we know that the average degree of the Apollonian network is given by

$$E\xi = \frac{3^{\frac{\ln 3}{\ln 2}-1} 2^{\frac{\ln 3}{\ln 2}}}{(2^{\frac{\ln 3}{\ln 2}-1}-1)3^{\frac{\ln 3}{\ln 2}-1}} = 6 \tag{17}$$

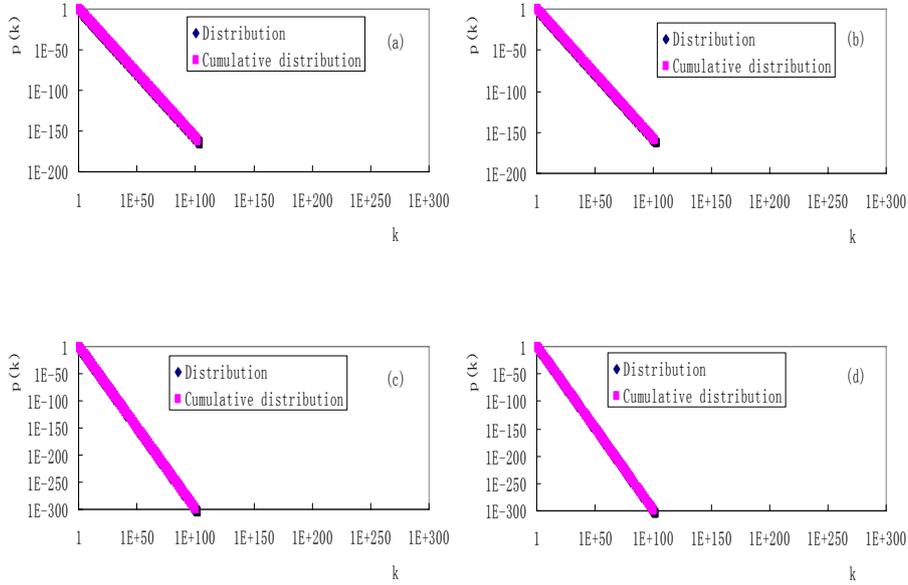

**FIG.1.** Numerical simulations of the power-law distribution with geometrically increasing range and its cumulative distribution. c=3. (a) m=1.2, r=ln(3)/ln(2)-1；(b) m=2, r=ln(3)/ln(2)-1；(c) m=1.2, r=2；(d) m=2, r=2.